\documentclass[preprintnumbers,amsmath,amssymb]{revtex4}
\usepackage{graphicx}
\usepackage{dcolumn}
\usepackage{bm}
\usepackage{subfigure}
\usepackage{epsfig}
\usepackage{float}
\usepackage{tensind}
\tensordelimiter{?}

\usepackage{mathrsfs}

\begin{document}

\newcommand{\bs}{\boldsymbol}
\newcommand\eqn[1]{\begin{eqnarray} #1 \end{eqnarray}}
\newcommand\vect[1]{\boldsymbol{#1}}
\newcommand\mat[1]{\mathsf{#1}}
\newcommand\trans{^\mathsf{T}}

\newcommand\braket[3]{\left<#1\,\right|#2\left|\,#3\right>}
\newcommand\ket[1]{\left|\,#1\right>}
\newcommand\bra[1]{\left<#1\,\right|}
\newcommand\braketsc[2]{\left<#1\,|\,#2\right>}
\newcommand\eg{\emph{e.g. }}
\newcommand\ie{\emph{i.e. }}
\newcommand\nn{\mathrm{N}_2}
\newcommand\hh{\mathrm{H}_2}
\newcommand\methane{\mathrm{CH}_4}
\newcommand\ris{\{\vect{r}_i\}}
\newcommand\rris{\{\vect{R}_i\}}
\newcommand\dris{d\vect{r}_i^N}
\newcommand\x{\vect{x}}
\newcommand\xis{\{\vect{x}_i\}}
\newcommand\xxis{\{\vect{X}_i\}}
\newcommand\dxis{d\vect{x}_i^N}

\title{On the Geometrization of Quantum Mechanics}
\author{Ivano Tavernelli} 
\thanks{Current address: IBM Research GmbH, Zurich Research Laboratory, 8803 R\"uschlikon.}
\email{ita@zurich.ibm.com}
 \affiliation{
 IBM Research GmbH, Zurich Research Laboratory, 8803 R\"uschlikon.}
\date{October 21, 2015}

\begin{abstract}
Nonrelativistic quantum mechanics is commonly formulated in terms of wavefunctions (probability amplitudes) obeying the static and the time-dependent Schr\"odinger equations (SE). 
Despite the success of this representation of the quantum world a wave-particle duality concept is required to reconcile the theory with observations (experimental measurements).
A first solution to this dichotomy was introduced in the de Broglie-Bohm theory according to which a pilot wave (solution of the SE) is guiding the evolution of  particle trajectories.
Here, I propose a geometrization of quantum mechanics that describes the time evolution of particles as geodesic lines in a curved  space, whose curvature is induced by the quantum potential.
This formulation allows therefore the incorporation of all quantum effects into the geometry of space-time, as it is the case for gravitation in the general relativity.
 \end{abstract}

\maketitle

\section{Introduction}

Despite the enormous success of the theory of general relativity, not many
attempts were made to apply the same geometrical approach to the description of the
dynamics induced by other (fundamental) forces such as electrostatic in 
quantum mechanics~\cite{Feynman_grav}.
One of the main reasons for that, is the triumph of the wavefunction interpretation 
of the quantum world, and in particular of quantum electrodynamics
and its description of the fundamental interactions as an exchange of virtual 
boson particles governed by the uncertainty principle. 
An important aspect that makes quantum mechanics hard to reconcile with a geometric description of the 
dynamics in terms of trajectories in configuration space is quantum correlation. 
In fact, wave mechanics as described by the Schr\"odinger equation (SE) offers a natural framework for the interpretation of coherence and decoherence effects and other peculiar properties of quantum mechanics like for instance the state superposition principle~\cite{comment_1}.
All these phenomena seem to rule out the possibility to include determinism in the description of quantum dynamics.
Nonetheless, a first attempt in this direction 
was done by De Broglie and Bohm~\cite{deBroglie26b, Bohm1952a,Bohm1952};
they introduced the concept of material point trajectories driven by a pilote wave that evolves according to the 
time-dependent Schr\"odinger equation for the system wave function 
$\psi(x_1, \dots, x_n, t)=\psi(\bs{x},t)$, with $x_i \in \mathbb{R}^3$.
In this picture, Bohmian trajectories follow the flux lines of the quantum probability current,
$\mathbf{j}(\bs{x})=(\mathbf{j}_1(\bs{x}), \dots, \mathbf{j}_n(\bs{x}))$ with 
$\mathbf{j}_i(\bs x,t)=\hbar/m^{-1} \Im (\psi^*(\bs{x},t) \nabla_i \psi(\bs{x},t))$,
which depends on the position of all particles in the system $(x_1, \dots, x_n)$ and introduces therefore nonlocality~\cite{endnote_nonlocality} into the dynamics.
Equivalently,  the same dynamics can be reformulated in terms of configuration space trajectories following a Newton's-like evolution (second order in time~\cite{endnote1}
in which the quantum potential $Q(\bs{x},t)=-(\hbar^2/2 m) \nabla^2A(\bs{x},t)/A(\bs{x},t)$ with $A(\bs x,t )=(\psi^*(\bs{x},t)\psi(\bs{x},t))^{1/2}$
is added to the classical potential and contributes to the forces driving the system~\cite{hollandbook, duerr92}.
In the Bohmian mechanics~\cite{Bohm1952a} the time evolution of a quantum system is described by a single trajectory in phase space that is driven by the nonlocal quantum potential $Q(\bs{x},t)$, which is also a function of the entire configuration space and time.

In this work, I show that the effect of the Bohmian quantum potential can be absorbed into the geometry of the configuration space 
giving rise to a quantum dynamics  described by deterministic trajectories evolving in a curved configuration space.
The quantum potential is defined in the $3 n$ dimensional configuration space, which makes this theory \textit{from-the-start} a many-body formulation of the dynamics (for $n\geq2$, where $n$ is the number of particles in the system of interest) and induces nonlocality in the dynamics associated to all constituent particles.
The geometrization of the physical space is performed in a Finsler differential manifold~\cite{Rund59}, which is a nontrivial generalization of Riemann space in which the metric tensor depends from both positions and momenta.
Different geometrization schemes based on a generalization of Riemann geometry were already proposed in the past, starting from the work of Weyl on electromagnetism~\cite{weyl1918}. 
Interesting extensions to quantum mechanics were more recently proposed by Novello and coworkers using a `Weyl integrable space' (Q-wis) geometry~\cite{novello2011} obtained from a variational principle.
In this work, we propose a geometrization scheme in which the metric tensor is derived from the quantum potential instead of being inferred; the price to pay is the extension from Riemann (and Q-wis geometry) to Finsler spaces.
Finally, it is also interesting to mention the interesting approach by Ootsuka and Tanaka that deals with Feynman path-integrals 
in an extended configuration space endowed with a Finsler metric~\cite{ootsuka2010}. 

\section{Theory}
\subsection{Finsler geometry} 

For non-conservative systems described by time-dependent Lagrangians (and therefore time-dependent potentials like the quantum potential $Q(\bs x,t)$), the dynamics can be reformulated by means of a homogeneous formalism in an extended configuration space of dimension ($3n+1$) where the additional variable corresponds to the time~\cite{Schouten,endnote0}.
Defining $M$ the configuration space manifold and $TM$ the corresponding tangent space, 
if $\mathcal{L}: \mathbb{R} \times TM \rightarrow \mathbb{R}$ is the time-dependent Lagrangian function, then a corresponding generalized homogeneous Lagrangian $\Lambda: T(\mathbb{R} \times M) \rightarrow \mathbb{R}$ can be defined by
\begin{equation} \label{eq:defLambda}
\Lambda(t, {\bs{x}}, \dot{t}, {\dot{\bs{x}}}) \equiv  \dot{t} \mathcal{L}(t, {\bs{x}}, {\bs{x}}') = \dot{t} \mathcal{L}(t, {\bs{x}}, {\dot{\bs{x}}}/  \dot{t})
\end{equation}
with $(x_{i})'=d x_i/dt$,  $\dot{x}^i=d x_i/d\tau$, ($i=1,\dots, n$).
In this formalism a new parameter $\tau(t)$ has been introduced to trace the progress of the system in the extended configuration spaces called the \textit{event} space.
As shown in Appendix A, the Euler-Lagrange equations corresponding to the Lagrangian $\Lambda(t, {\bs{x}}, \dot{t}, {\dot{\bs{x}}})$ are invariant with respect to any regular transformation of the parameter $\tau(t)$, and for the sub manifold obtained by setting $\dot{t}= 1$ they reproduce the equations of motion for  $\mathcal{L}(t, {\bs{x}}, {\dot{\bs{x}}})$.
In the extended configuration space, time is therefore risen to the rank of an additional generalized coordinate: $t=q_0 \in \mathbb R$, and the dynamics is described in the $(3n+1)$-dimensional manifold (space of events) span by the generalised coordinates and velocities $\{ \{q^{\alpha}\}_{\alpha=0}^{n}, \{\dot{q}^{\alpha}\}_{\alpha=0}^{n} \}\equiv \{\bs q, \dot{\bs q} \}$. 
In the following, I will relabel the coordinates of the canonical configuration space as $\{t, \{{x}_i\}_{i=1}^{n}\}$,  and I will use $\{q^{\alpha}\}_{\alpha=0}^{n}$ for the extended configuration space according to the identification: $q^0=t$, $\dot{q}^0=dt/d\tau$, ${q}^i={x}_i$ ($i=1,\dots, n$) with $x_i \in \mathbb{R}^3$ and $q^i \in \mathbb{R}^3$; the velocities are defined as $\{\dot{q}^{\alpha}=d q_{\alpha}/ d \tau \}_{\alpha=0}^{n}$, and $\{({x}_i)'=d {x}_{i}/ d t \}_{i=1}^{n}$.
For a Lagrangian with time-dependent potential $V({\bs{x}},t)$ of the form
\begin{equation} \label{eq:lagrangian_vtx}
\mathcal{L}(t, {\bs{x}}, {\bs{x}}') =  T({\bs{x}}') -V({\bs{x}},t)
\end{equation}
we have therefore
\begin{equation}
\Lambda({\bs{q}},\dot{{\bs{q}}}) = \mathcal{T} (\{\dot{q}^i\}^n_{i=1})/ \dot{q}^0 - V({\bs q}) \, \dot{q}^0
\label{eq_def_Lambda}
\end{equation}
with 
$T({\bs{x}}') = \frac{1}{2} \sum_{i=1}^{n} m {(x_i)'} {(x_i)'}$ 
and $ \mathcal{T} (\{\dot{q}^i\}^n_{i=1}) =  \frac{1}{2}  \sum_{i=1}^{n} m (\dot{q}^i)^2 = T( {\bs{x}}') \, \, (\dot{q}^0)^2$ .

In the case of conservative systems with time-independent potentials,  using Jacobi's theorem it is possible to describe a dynamics in a given potential with a geodesic motion in a Riemannian manifold with suited metric~\cite{Abraham_Marsden}.
In the non-conservative case, where the dynamics is described by a homogeneous Lagrangian $\Lambda({\bs{q}},\dot{{\bs{q}}})$ in Eq.~\eqref{eq_def_Lambda}, geometrization is obtained in the framework of Finsler's spaces~\cite{Rund59}.
In a Finsler space $M^{(3n+1)}$ with coordinates $\bs{q}=(q^0, \dots, q^{n})$ the line element between two adjacent points in space is given by
\begin{equation} \label{eq:Fxxdot}
ds=\left( {g}_{\alpha \beta}(\bs{q}, \dot{\bs{q}}) dq^{\alpha} dq^{\beta} \right)^{1/2} = \Lambda(\bs{q}, \dot{\bs{q}}) d \tau \, .
\end{equation}
(Einstein's summation is assumed throughout the paper).
The main difference with a Riemannian space is that the metric tensor ${g}_{\alpha \beta}(\bs{q}, \dot{\bs{q}})$ depends also on velocities of the tangent space $TM^{(3n+1)}_{\bs{q}}$.
The Finsler metric in Eq.~\eqref{eq:Fxxdot} defines a dynamical system through the minimisation of the `action' functional $I(\gamma)=\int_{\tau_1}^{\tau_2} \Lambda(\bs{q}, \dot{\bs{q}})\, d\tau$ for a path $\gamma$ and given initial and final conditions ($\gamma(\tau_1)$ and $\gamma(\tau_2)$), 
when the  following three conditions are fulfilled~\cite{Caratheodory} 
(i) positive homogeneity of degree one in the second argument, $\Lambda(\bs{q},k \dot{\bs{q}})= k \Lambda(\bs{q}, \dot{\bs{q}}), k>0 $, 
(ii) $\Lambda(\bs{q}, \dot{\bs{q}})\neq 0, \forall \dot{\bs{q}}\neq0$, and 
(iii) $\frac{\partial^2 \Lambda^2(\bs{q}, \dot{\bs{q}})}{\partial \dot{q}^{\alpha} \partial \dot{q}^{\beta}} \xi^{\alpha} \xi^{\beta} > 0, \forall \bs{\xi}\neq \lambda \dot{\bs{q}}$. 

The next step consists in the geometrization of the quantum dynamics using a Finsler's metric derived from the time-dependent quantum Bohmian potential.
This article is organised as following.
First I derive a trajectory-based solution of the quantum dynamics starting from a general Lagrangian density for a complex scalar field $\phi(\bs x,t)$. 
In a second step, I will introduce the geometrization of the resulting dynamics  in a Finsler manifold~\cite{Rund59} defined on the extended configuration space of dimension $(3n+1)$. 
The equations of motion are given by geodesic curves on a curved space-time manifold whose metric tensor is derived from the quantum potential, which in turn is a functional of the system wavefunction, $\phi(\bs{x},t)$.

\subsection{Matter field and the trajectory representation of quantum dynamics.} 

Quantum dynamics can be formulated using from a canonical formalism in which the system wavefunction is treated as a classical complex scalar field $\phi(\bs x,t)$ associated to the Lagrangian density~\cite{schleich2013schrodinger,Takabayasi}
(using: $\{{q}^i\}^n_{i=1} = \{x_i\}_{i=1}^{n} \equiv \bs x$)
\begin{equation}
\mathscr{L}=\frac{i\hbar}{2} (\phi^*\dot{\phi}-\dot{\phi}^*\phi)
-\frac{\hbar^2}{2m} \partial_i \phi \partial_i \phi^* - V \phi\phi^* \, .
\label{eq:field3}
\end{equation}
Applying the principle of least-action to this Lagrangian density one obtained back the time dependent Schr\"odinger equation.
The same formalism can also be generalised to the case of the Dirac equation~\cite{Lurie}.
Here we are interested in deriving a trajectory representation of the dynamics associated to the Lagrangian in Eq.~\ref{eq:field3}. 
Following~\cite{Takabayasi, hollandbook, Misnerbook}, we start from the field conservation law 
\begin{equation}
\partial_{\alpha} ?T^{\alpha}^{}_{\beta}?=\frac{\partial \mathscr{L}}{\partial q_{\beta}} \, ,
\label{eq:field1}
\end{equation}
of the stress-energy-momentum tensor
\begin{equation}
?T^{\alpha}^{}_{\beta}?=
-\left[
\frac{\partial \mathscr{L}}{\partial(\partial_{\alpha} \rho)} \partial_{\beta} \rho + 
\frac{\partial \mathscr{L}}{\partial(\partial_{\alpha} S)} \partial_{\beta} S
\right] + ?\delta^{\alpha}^{}_{\beta}? \mathscr{L} \, ,
\label{eq:field2}
\end{equation}
where $S(\bs x,t)/  \hbar$ is the phase of $\phi(\bs x,t) = A(\bs x,t) e^{i S(\bs x,t)/ \hbar}$ and $\rho(\bs x,t)=A^2(\bs x,t)$.
For the energy density, $?T^{0}^{}_{0}?$, 
\begin{equation}
\frac{\partial ?T^0^{}_0?(\bs x,t)}{\partial t} + ?T^{i}^{}_{0,i}?(\bs x,t) = - \rho \frac{\partial V(\bs x)}{\partial t} \, ,
\label{eq:field1.1}
\end{equation}
($?T^{i}^{}_{0,i}?=\partial ?T^{i}^{}_{0}?/\partial x_{i}$) the conservation law can be reformulated in terms of the coupled differential equations for the amplitude and the phase of $\phi(\bs x,t)$ (Appendix E) 
\begin{align}
 \frac{\partial S(\bs{x},t)}{\partial t} &= - \left(V(\bs{x}) + Q(\bs{x},t) \right) -\frac{1}{2 m}  (\nabla S(\bs{x},t))^2 \,  \label{prop3} \\
 \frac{\partial A(\bs{x},t)}{\partial t} &= - \frac{1}{2 m} \big(2 \nabla A(\bs{x},t)   \nabla S(\bs{x},t) + A(\bs{x},t) \Delta S(\bs{x},t) \big)  \label{prop4}
\end{align}
where $Q(\bs x,t)=-\frac{\hbar^2}{2m}\frac{\nabla^2 A(\bs x,t)}{A(\bs x,t)}$ is the quantum potential.
The dynamics described in Eq.~\eqref{prop3} is equivalent to the Bohmian trajectory dynamics 
$ \bs x'(t)= v^{\phi}(\bs{x},t) $
for the vector field~\cite{Bohm1952} (Appendix D)
\begin{equation} \label{theo1}
v^{\phi}(\bs{x},t)=\frac{\hbar}{m} \Im \frac{\nabla_{\bs{x}} \phi(\bs{x},t)}{\phi(\bs{x},t)} \, .
\end{equation} 
Alternatively, one can identify Eq.~\eqref{prop3} with the Hamilton-Jacobi equation for the `classical' dynamics in the potential $V(\bs x)+Q(\bs{x},t) $. Its solution by characteristics, for given initial conditions, corresponds to the trajectory solution of the 
\textit{Newton-like} equation of motion~\cite{nabdy2,nabdy3} 
\begin{equation} \label{theo2}
m \frac{d}{dt} \bs{x}'(t) =- \nabla_{\bs{x}} (V(\bs{x})+Q(\bs{x},t)) \, 
\end{equation}
with corresponding Lagrangian
\begin{equation} \label{theo3}
\mathcal{L}(\bs{x},{\bs{x}}',t)=\mathcal{T}(\bs{x}')- (V(\bs{x})+ Q(\bs{x},t))) \, . 
\end{equation} 
In Eq.~\eqref{theo2} $\bs{x}'$  stands for $\partial \bs{x}/\partial t$ and $d/dt=\partial/\partial t + \bs x'(t)\cdot\nabla$.
However, a word of caution is recommended here, since the equivalence of the first- and second-order formulations of 
the dynamics (Eqs.~\eqref{theo1} and ~\eqref{theo2})  is still debated~\cite{valentini2008}.

The next step consists in the geometrization of the quantum dynamics obtained by absorbing the quantum potential $Q(\bs{x},t)$ into the metric tensor of the Finsler space. 
This is described in the following \textit{proposition}.

\subsection{Proposition} 

Consider a system 
of particles with coordinates $q^i \in \mathbb{R}^3$, $i=1,\dots, n$.
The dynamics takes place in the extended configuration space 
with $q^{0}=t$, while the progress of the dynamics is measured in terms of a \textit{proper} time parameter, $\tau$.
For any given initial condition, the quantum dynamics associated to each configuration point follows a deterministic trajectory
in the curved $3 n+1$ dimensional space according to the geodesic equation
($\mu, \nu, \xi=0,\dots, 3n$)
\begin{equation} \label{prop1} 
\ddot{q}^{\mu}+?\Gamma^{\mu}^{}_{\nu \xi}?(\bs q,\dot{\bs q}) \dot{q}^{\nu} \dot{q}^{\xi} = - {g}^{\mu \nu} \partial V(\{q_i\}^n_{i=1}) /\partial q^\nu
\end{equation}
where 
$?\Gamma^{\mu}^{}_{\nu \xi}?(\bs q,\dot{\bs q}) =\frac{1}{2} {g}^{\mu\sigma}({g}_{\sigma\xi,\nu}+{g}_{\sigma\nu,\xi}-{g}_{\nu \xi,\sigma})$ are the generalized connections (with ${g}_{\sigma\nu,\xi}=\partial_{{\xi}}{g}_{\sigma\mu}\equiv \partial g_{\sigma \mu}/\partial q_{\xi}$), 
and the space metric ${g}_{\mu\nu}$ is given by (Appendix C)
\begin{equation} \label{prop2}
g_{\mu \nu}(\bs q,\dot{\bs q})=\frac{1}{2} \frac{\partial^2 \Lambda^2(\bs q,\dot{\bs q})}{\partial \dot{q}^{\mu} \partial \dot{q}^{\nu}} \,
\end{equation}
with 
\begin{equation} \label{prop2b}
\Lambda(\bs q,\dot{\bs q})=\mathcal{T} (\{\dot{q}^i\}^n_{i=1})/\dot{q}^0  - Q(\bs{q}) \dot{q}^0 \, 
\end{equation}
and $\mathcal{T} (\{\dot{q}^i\}^n_{i=1})=(1/2) m (\dot{q}^i)^2$ ($m$ is the particle mass).
In Eqs.~\eqref{prop1}-\eqref{prop2b} we use $\tau=s$, where $s$ is the arclength defined by $ds=\Lambda(\bs{q},\dot{\bs{q}}) d\tau$; 
$\bs{q}(t)$ and $\dot{\bs{q}}(t)$ are explicit functions of time, $V(\{{q}^i\}^n_{i=1})$ is the classical potential, and $ Q(\{{q}^{i} \}^n_{i =1},t)= Q(\bs{x},t)$
is the quantum potential. \\

\textit{Proof:} The proof is organised in two steps. \textit{(i)} Starting from the Lagrangian density in Eq.~\eqref{eq:field3} one can describe quantum dynamics 
using trajectories following the Newton-like equation of motion given in Eq.~\eqref{theo2}. Equivalently, the same result can be obtained following the derivation 
due to Bohm~\cite{Bohm1952a}. 
\textit{(ii)} The time-dependent Lagrangian associated to this dynamics (Eq.~\eqref{prop2b}), can be \textit{geometrized} using Finsler's approach 
outlined in Eqs.~\eqref{eq:lagrangian_vtx}-\eqref{eq:Fxxdot}, which leads to the geodesic dynamics given in Eq.~\eqref{prop1} (see also Appendices A and B). $\square$
\bigskip

In this proposition, a new description of quantum mechanics is proposed, in which the particle-wave duality is completely overcome. 
Matter is represented as point particles that move in a curved configuration space where the curvature is self-induced as well as generated, in a nonlocal fashion, by all other particles in the system; in other words, the metric tensor (and ultimately the curvature) depends on the coordinate of the particle of interest $q^i$, as well as from all other particles in the system, $q^{j}$, with $j\neq i$~\cite{comment_2}.
The space curvature generated by the  quantum potential evolves according to Eq.~\eqref{prop3} and, therefore, the wave-like nature of quantum mechanics is confined to the geometric description of the space and voided from any mechanical property; 
all quantum effects are now absorbed into the metric of the configuration space~\cite{comment_3}.
It is worth mentioning that the geometrization of the configuration space is unique in the sense that is
driven by the complex field $\phi(\bs{x},t)$, which is determined by the Lagrangian density 
Eq.~\eqref{eq:field3} or, 
equivalently, the system of equations in~\eqref{prop3} and ~\eqref{prop4}. 
The connection to the  Schr\"odinger is discussed in the next section.

\subsection{Relation to the Schr\"odinger dynamics} 

An ensemble of trajectories following the dynamics in Eq.~\eqref{prop1}  with properly chosen initial conditions describes  the wavefunction  dynamics of the time-dependent SE. 
In fact, when projected back to the $3n$ dimensional configuration space the trajectories defined in Eq.~\eqref{prop1} reproduce exactly the Bohmian trajectories. This is evident from the correspondence of the characteristics associated to Eq.~\eqref{prop3} with the Bohmian trajectories tangent to the vector field in Eq.~\eqref{theo1}. 
Therefore, when the initial configurations are sampled according to a given initial wavefunction density $|\phi(\bs x,0)|^2$, the geodesics (or the corresponding Bohmian trajectories) evolve these points such that at a later time their density, $\rho(\bs x, t)$, is 
consistent with the time-propagated wavefunction $\phi(\bs x,t)$.

\section{Application} 

\begin{figure}[h]
\begin{center}
\includegraphics[width=8.5 cm]{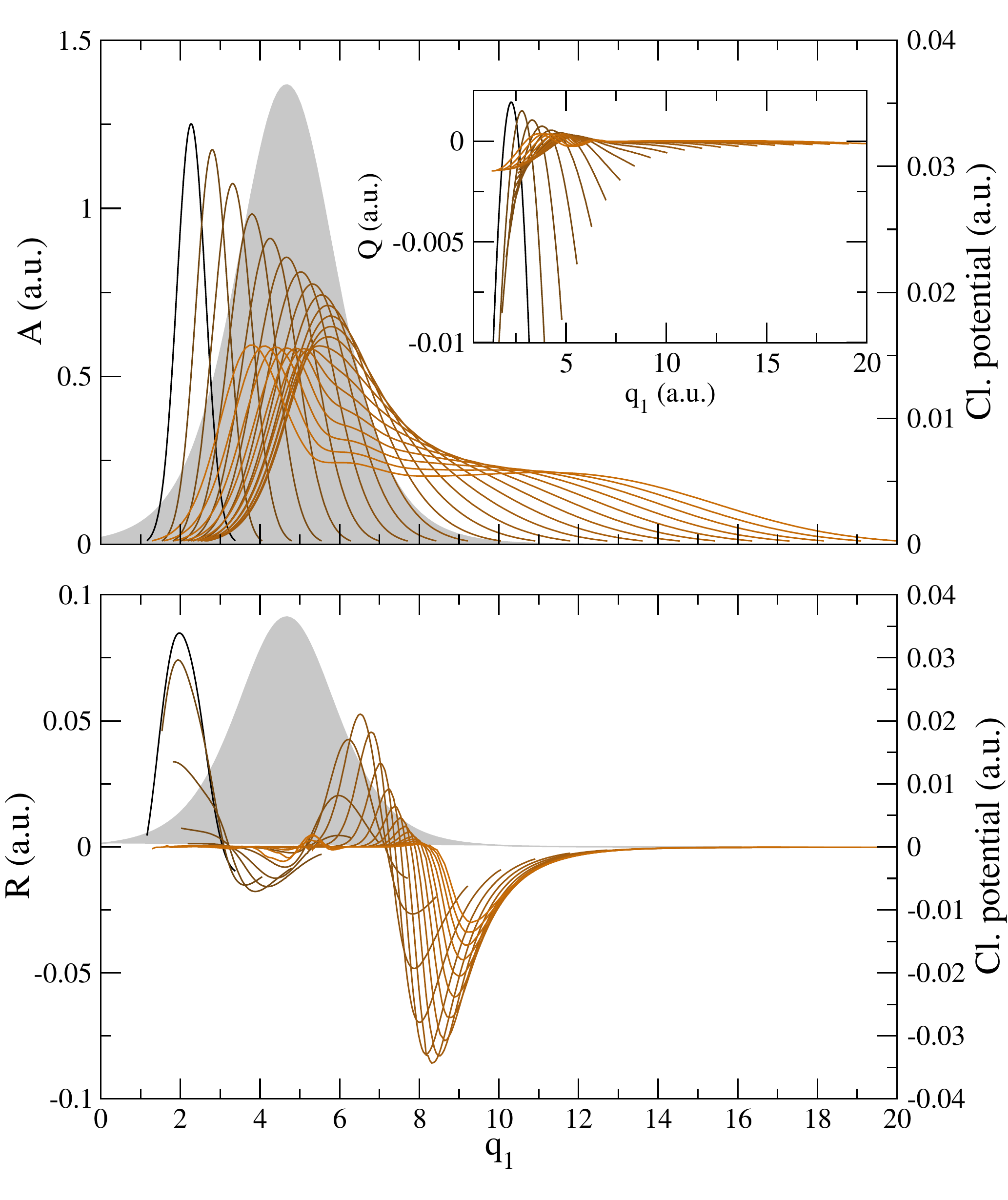}
\caption{Upper panel: Time evolution of the nuclear wavepacket. Snapshots are taken between time $t=50$ (dark brown) and $t=2100$ (light brown) in intervals of $100$. The inset shows the dynamics of the quantum potential in the same time interval.
Lower panel: Time evolution of the space curvature $R(\bs{q},\dot{\bs{q}})$ projected along the space coordinate $q_1$ (same color code as for the upper panel). The static classical potential (Eckart barrier) is shown in gray. All quantities are measured in atomic units.}
\label{Fig1}
\end{center}
\end{figure}

\begin{figure}[h]
\begin{center}
\includegraphics[width=8.5 cm]{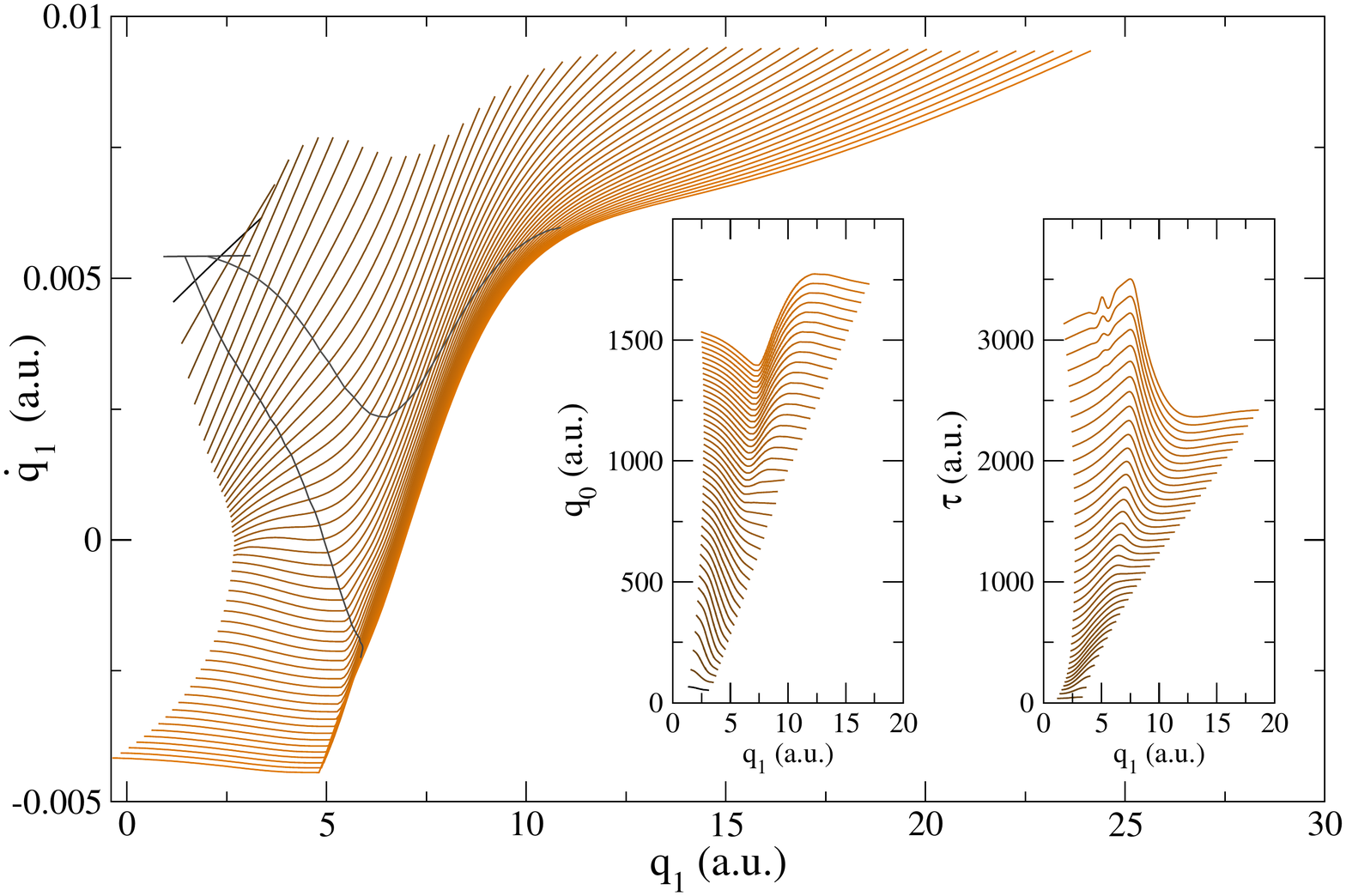}
\caption{Time evolution of the \textit{line fronts} defined by the variables $\{ q_{(i)}^1(t), \dot{q}_{(i)}^1(t)\}$ along the trajectories, $i=1,\dots, 200$ at regular time intervals of $50$ a.u. starting from $t=0$. Color code: brown scale from black (t=0 a.u.) to orange (t=2100 a.u.). Two trajectories are shown in black. The same dynamics projected on the $\{q^1,q^0\}$ and $\{q^1,\tau \}$ planes is shown in the two insets.}
\label{Fig2}
\end{center}
\end{figure}

I illustrate the implications of the dynamics described in the Proposition by means of a simple example: the scattering of a Gaussian wavepacket with an Eckart-type potential barrier in one dimension (Fig.~\ref{Fig1}). This problem is described in details in~\cite{kendrick2003new}.
In this case, the Finsler's space is described by an extended $2+2$ dimensional phase space with coordinates $\{ q^0=t, \dot{q}^0, q^1, \dot{q}^1\}$, while the progress of the dynamics is measured by the parameter $\tau$.
The wavepacket, $A(q^1,0)=(2\beta/\pi)^{1/4} e^{-\beta(q^1-q_c)^2} e^{ik(q^1-q_c)}$ ($\beta=4$, $k=10.8842$) is initialised at position $q_c=2$, while the potential, 
$V(q^1)=V_0 \, \text{sech}^2 (a (q^1-q_p))$, is fixed and centered at $q_{p}=7$. All quantities are given in atomic units (a.u.), $a=0.4$, $V_0=0.0365$.
An ensemble of 200 trajectories is used to describe the wavepacket dynamics. They are initially distributed according the the probability $A^2(q^1(0))$, while $\dot{q}^1(0)=1$, $q^0(0)=0$ and $\dot{q}^0(0)=1$ for all trajectories. 
The time evolution of the wavepacket is depicted in Fig.~\ref{Fig1} together with the corresponding snapshots of the quantum potential computed on the support of the wavepacket (inset of the upper panel), which is given by the values $q_1$ sampled by the 200 trajectories at a given value of $\tau$ (or time $t(\tau)$).
The space curvature is defined as ($(\bs{q},\dot{\bs{q}}) \equiv (q^0(t),\dot{q}^0(t),q^1(t),\dot{q}^1(t))$)
\begin{equation}
R(\bs{q},\dot{\bs{q}})= g^{\alpha \beta}(\bs{q},\dot{\bs{q}}) R_{\alpha \beta} (\bs{q},\dot{\bs{q}})
\end{equation}
where $R_{\alpha \beta}=?R^{\gamma}^{}_{\alpha\beta\gamma}?$ is the Ricci tensor and 
 \begin{align} \label{Eq_riemann_curv}
?R^{\gamma}^{}_{\alpha\beta\gamma}?&(\bs{q},\dot{\bs{q}}) = ?\Gamma^{\delta}^{}_{\alpha \gamma, \beta}? (\bs{q},\dot{\bs{q}}) -
?\Gamma^{\delta}^{}_{\alpha \beta, \gamma}? (\bs{q},\dot{\bs{q}})+ \notag \\
& ?\Gamma^{\lambda}^{}_{\alpha \gamma}? (\bs{q},\dot{\bs{q}}) ?\Gamma^{\delta}^{}_{\lambda \beta}? (\bs{q},\dot{\bs{q}})-
?\Gamma^{\lambda}^{}_{\alpha \beta}? (\bs{q},\dot{\bs{q}}) ?\Gamma^{\delta}^{}_{\lambda \gamma}? (\bs{q},\dot{\bs{q}})
\end{align}
is the Riemann curvature tensor with $?\Gamma^{\delta}^{}_{\alpha \gamma, \beta}? = \partial ?\Gamma^{\delta}^{}_{\alpha \gamma}?/\partial q^{\beta}$.
The lower panel of Fig.~\ref{Fig1} shows the value of the curvature, $R(q_{(i)}^1(\tau_j);q_{(i)}^0(\tau_j),\dot{q}_{(i)}^0(\tau_j),\dot{q}_{(i)}^1(\tau_j))$, on the support of the wavepacket, sampled by the trajectories labeled with $i$ ($i=1, \dots, 200$), at different 
values of $t$ ranging from $t=50$ to $t=2100$ in intervals of $100$.
Initially ($t<400$), the curvature is positive for all points (dark brown lines), while at later times we observe regions of positive and negative curvature. Towards the end of the simulation the curvature become strongly negative for all points at the right-hand side of the potential barrier.
This becomes more evident if we observe the dynamics in the plane defined by the coordinates $q^1$ and $\dot{q}^1$. 
Fig.~\ref{Fig2} shows the time evolution of the system as a series of \textit{line fronts} in the  $q^1$, $\dot{q}^1$ plane; 
each line is defined by the points $\{q^1_{(i)}, \dot{q}^1_{(i)} \}$ at a given time $t$, where the index $i$ runs over the ensemble of trajectories. 
The dynamics of two representative trajectories projected on the same plane (black lines) shows a first `convergent' behavior (signature of a positive curvature) followed by a diverging behavior (negative curvature). 
Of particular interest is the analysis of the time dilation effect. 
The right inset in Fig.~\ref{Fig2} shows the line fronts in the $(q^1,q^0)$ plane, each line corresponding to a fixed value of the `reference time'  $\tau$. We observe that the local time, $q^0(\tau)$, runs faster for the points on the right-hand side of the wavepacket (larger gaps between the line) than for the points on the left-hand side that are recoiled by the potential.

\section{Conclusions}

In conclusion, in this article I present a geometrization of quantum dynamics according to which all non-classical quantum effects are included in the  geometry of an extended phase space (that includes time) of dimension $6n+2$. The particles evolve along geodesic lines in a curved Finsler manifold modeled by the quantum potential. 
In this framework, the particle-wave duality of conventional quantum mechanics is replaced by deterministic particle dynamics evolving on a curved manifold, where the curvature is derived from the \textit{nonlocal} quantum potential that depends on the position of all particles in the system (independently from their reciprocal distance and interaction).
This picture of quantum dynamics is well suited for the understanding of nonlocality in quantum mechanics including, for instance, 
the self-interference of single particles (electrons or molecules) on a double slit~\cite{PhysRevLett.87.160401, PhysRevLett.103.263601,juffmann2012real}, and 
 the interpretation of the Mach-Zehnder's interferometer~\cite{Philippidis1979,Rarity1990}, for which particle-wave duality is required according to the classical interpretation of quantum mechanics.
In the first situation, the interference pattern results from the nonlocality of the quantum potential, $Q({\bs q}_\text{slit}, \bs{q}_{\text{mol}},t)$, which entangles the single particle dynamics, $\bs{q}_{\text{mol}}(t)$, with the one of the device (double slit).
Mach-Zehnder's interference at a distance is made possible by the fact that quantum information is not just carried by the particle dynamics but also includes a component  from the propagation of the configuration space curvature~\cite{endnote3,comment_4}.
This interpretation of quantum dynamics offers in addition a very simple solution to the measurement problem in quantum mechanics; in fact, like in Bohmian mechanics~\cite{Bohm1952a}, the trajectories describe deterministic paths in space and the transition to the classical world does not require  the concept of wavefunction collapse.
The wavefunction character of the quantum state and its dynamics is now confined  to the characterization of the curvature of the configuration space of each constituent particle, while the system evolves along deterministic geodesic curves. 
Finally, the geometrization of quantum mechanics offers a clear opportunity for the unification with Einstein's theory of general relativity.

\begin{acknowledgments}
The author acknowledges  Giovanni Ciccotti for useful discussions and Basile Curchod for his support with the numerical calculations. This work was made possible thanks to the financial support of the Swiss National Science Foundation (SNF) through grant No. 200021-146396. 
\end{acknowledgments}

\appendix
\numberwithin{equation}{section}
\section{Equivalence of the dynamics described by $\Lambda(\bs{q},\dot{\bs{q}})$ and by $\mathcal{L}(\bs{x},\dot{\bs{x}},t)$}

The demonstration of the equivalence  of the dynamics generated by $\Lambda(\{q^{\alpha}\}_{\alpha=0}^{n},\{\dot{q}^{\alpha}\}_{\alpha=0}^n)=\Lambda(\bs{q},\dot{\bs{q}})$ and $\mathcal{L}(t,\{x_{i}\}_{i=1}^n,\{({x}_i)'\}_{i=1}^n)$ is given in two parts. 
I will use the following notation: $q^0=t$, $\dot{q}^0=dt/d\tau$, ${q}^i={x}_i$ ($i=1,\dots, n$) with $x_i \in \mathbb{R}^3$ and $q^i \in \mathbb{R}^3$; the velocities are defined as $\{\dot{q}^{\alpha}=d q^{\alpha}/ d \tau \}_{\alpha=0}^{n}$, and $\{x'_i=d {x}_{i}/ d t \}_{i=1}^{n}$.
In the first part, I show that the Lagrange dynamics in the extended ($3n+1$) configuration space (governed by the Lagrangian $\Lambda(\{q^{\alpha}\}^n_{\alpha=0} ,\{\dot{q}^{\alpha}\}^n_{\alpha=0} $) is equivalent to the \textit{standard} Lagrange formulation of dynamics in the $n$-dimensional configuration space (governed by the Lagrangian $\mathcal{L}(t,\{q^i\}^n_{i=1},\{\dot{q}^i\}^n_{i=1})$. 
The dependence on $\{q^{\alpha}\}^n_{\alpha=0}$ and $\{q^i\}^n_{i=1}$ indicates a dependence of the full ($3n+1$) and ($3n$) dimensional manifolds, respectively. 
In the following I will drop the indexes when referring to the entire vectors and use them only for the components:
\begin{equation}
 \{x_{i}\}_{i=1}^n = \{q^{i}\}_{i=1}^n \equiv \bs x, \,   
 \{(x_{i})'\}_{i=1}^n  \equiv \bs x', \, 
 \{q^{\alpha}\}_{\alpha=0}^n \equiv \bs q, \, \text{and} \,
 \{\dot{q}^{\alpha}\}_{\alpha=0}^n \equiv \bs \dot{\bs q} \, .
 \end{equation}
In the second step, I show that the Lagrange dynamics in the extended ($3n+1$) configuration space can be formulated as a geodesic motion in a curved manifold. 

The addition of a classical time-independent potential $V(\bs x)$ will also be considered at the end of this section.
\bigskip

\paragraph*{Step 1.} 
The Lagrange system obtained from the minimisation of the of the action functional $I(\gamma)=\int_{\tau_1}^{\tau_2} \Lambda(\bs q, \dot{\bs q})\, d\tau$,
\begin{equation} \label{eq_Lambda}
\frac{d}{d\tau} \left(\frac{\partial \Lambda(\bs q, \dot{\bs q})}{\partial \dot{q}^{\alpha}} \right) - \frac{\partial \Lambda(\bs q, \dot{\bs q})}{\partial {q}^{\alpha}} =0 \, , \quad \quad (\alpha=0,\dots,n)
\end{equation} 
consists of $3n+1$ equations among which only $3n$ are independent and correspond to the equations of motion associated to $\mathcal{L}(\bs q,\dot{\bs q})$.
To show the linear dependence of the $3n+1$ Lagrange equations, we first make use of the homogeneity condition of degree one in the velocity ($\Lambda(\bs q,k \dot{\bs q})= k \Lambda(\bs q,\dot{\bs q}), k>0 $), which according to Euler's theorem gives (H. Rund, Ref.~\cite{Rund59}, pp. 3-4)
\begin{align}
\frac{\partial \Lambda(\bs q, \dot{\bs q})}{\partial \dot{q}^{\alpha}}   \dot{q}^{\alpha} &= \Lambda(\bs q, \dot{\bs q})  \label{eq_euler1}\\
\frac{\partial^2 \Lambda(\bs q, \dot{\bs q})}{\partial \dot{q}^{\alpha} \partial \dot{q}^{\beta}}   \dot{q}^{\alpha} &= 0 \label{eq_euler2} \, ,
\end{align} 
with (see definitions in the main text)
\begin{equation}
\Lambda(\bs q,\dot{\bs q})=\mathcal{T}(\dot{\bs q})/\dot{q}^0 -Q(\bs x,t) \dot{q}^0 \, .
\end{equation} 
The following relation can therefore be established
\begin{align}
\sum_{\alpha} \dot{q}^{\alpha} 
\left(
\frac{d}{d\tau} \left(\frac{\partial \Lambda(\bs q, \dot{\bs q})}{\partial \dot{q}^{\alpha}} \right) - \frac{\partial \Lambda(\bs q, \dot{\bs q})}{\partial {q}^{\alpha}} 
\right) 
&=  \frac{d}{d\tau} \sum_{\alpha} \left( \dot{q}^{\alpha} \frac{\partial \Lambda(\bs q, \dot{\bs q})}{\partial {q}^{\alpha}} \right) 
- \sum_{\alpha} \left( \ddot{q}^{\alpha}  \frac{\partial \Lambda(\bs q, \dot{\bs q})}{\partial \dot{q}^{\alpha}} + \dot{q}^{\alpha} \frac{\partial \Lambda(\bs q, \dot{\bs q})}{\partial {q}^{\alpha}}  \right) \notag \\
&= \frac{d}{d \tau} \Lambda(\bs q, \dot{\bs q}) - \frac{d}{d \tau} \Lambda(\bs q, \dot{\bs q})  =0 \, .
\end{align} 
The system in Eq.~\eqref{eq_Lambda} is therefore equivalent to the $3 n$ \textit{standard} Lagrange equations
\begin{equation} \label{eq_L}
\frac{d}{dt} \left(\frac{\partial \mathcal{L}(t, \bs x, {\bs x}')}{\partial x'_{i}} \right) - \frac{\partial \mathcal{L}(t, \bs x, {\bs x}')}{\partial {x}_{i}} =0 \, , \quad \quad (i=1,\dots,n) \, .
\end{equation} 
while the extra equation sets the freedom for the choice of $\tau(q_0)$.
To show this, we take the choice $\tau=t$ and therefore $\dot{q}^0=1$.  
Substituting 
\begin{equation}
\Lambda(\bs q,\dot{\bs q}) = 
\left( T(\bs x')  - Q(\{q_i\}_{i=1}^n,t) \right) \, \dot{q}^0 = \mathcal{L}(t, \bs x, {\bs x}')
\end{equation} 
into Eq.~\eqref{eq_Lambda} we obtained exactly Eq.~\eqref{eq_L} (for $i=1,\dots,n$).

\bigskip

\paragraph*{Step 2.} The demonstration of the correspondence between the Euler-Lagrange equations for $\Lambda(\bs q, \dot{\bs q})$ and the geodesic equation in the curved manifold characterised by the metric tensor $g_{\alpha \beta}(\bs q,\dot{\bs q})=\frac{1}{2} \frac{\partial^2 \Lambda^2(\bs q,\dot{\bs q})}{\partial \dot{q}^{\alpha} \partial \dot{q}^{\beta}}$ follows closely the one given by H. Rund (Ref.~\cite{Rund59}, Chapter II, paragraph 2).

Starting from the Euler-Lagrange equations formulated in the arc-length parameter $s$ (for which $\Lambda(\bs q,\dot{\bs q})=1$ along the path)~\footnote{
The arc-length is defined by 
$ds = \Lambda(\bs q,\dot{\bs q}) d\tau $.
Setting $\tau=s$ one has $\frac{ds}{d\tau}=1=\Lambda(\bs q,\dot{\bs q})$.},
\begin{equation}
\frac{d}{ds} \left( \frac{\partial \Lambda(\bs q, \dot{\bs q})}{\partial \dot{q}^{\alpha}}\right)- \frac{\partial \Lambda(\bs q, \dot{\bs q})}{\partial q^{\alpha}}  = 0 \, , \quad  (\alpha=0, \dots n)
\end{equation} 
with the definition
\begin{equation}
g_{\alpha \beta} (\bs q, \dot{\bs q}) =\frac{1}{2} \frac{\partial^2 \Lambda^2(\bs q, \dot{\bs q})}{\partial \dot{q}^{\alpha} \partial \dot{q}^{\beta}}
\end{equation} 
together with the relation
\begin{equation}
\frac{1}{2} \frac{\partial^2 \Lambda^2(\bs q, \dot{\bs q})}{\partial \dot{q}^{\alpha} \partial \dot{q}^{\beta}} \dot{q}_{\beta} = \Lambda(\bs q, \dot{\bs q}) \frac{\partial \Lambda(\bs q, \dot{\bs q})}{\partial \dot{q}^{\alpha}} 
\label{Eq:deriv1}
\end{equation} 
we get 
\begin{equation}
\frac{\partial \Lambda(\bs q, \dot{\bs q})}{\partial q^{\gamma}} = \frac{d}{d s} (g_{\gamma \beta} (\bs q, \dot{\bs q}) \dot{q}^{\beta}) \,.
\label{eq_rhs}
\end{equation} 

The derivation of Eq.~\eqref{Eq:deriv1} is given at the end of this Appendix.

\vspace{0.5cm}
In view of the homogeneity condition of $\Lambda(\bs q, \dot{\bs q})$, ($\Lambda(\bs q, k \dot{\bs q})=k \Lambda(\bs q, \dot{\bs q}) $), $\Lambda^2(\bs q, \dot{\bs q})$ is positively homogeneous of second degree in the $\dot{ q}^{\alpha}$ and therefore from $g_{\alpha \beta}(\bs q,\dot{\bs q})=\frac{1}{2} \frac{\partial^2 \Lambda^2(\bs q,\dot{\bs q})}{\partial \dot{q}^{\alpha} \partial \dot{q}^{\beta}}$ it follows
\begin{equation}
\Lambda(\bs q,\dot{\bs q})=\sqrt{g_{\alpha\beta}(\bs q, \dot{\bs q}) \dot{q}^{\alpha} \dot{q}^{\beta}} \, .
\end{equation}

The left hand side of Eq.~\eqref{eq_rhs} can be rewritten as 
\begin{align}
\frac{\partial}{\partial q^{\gamma}} \Lambda(\bs q, \dot{\bs q}) &=\frac{\partial}{\partial q^{\gamma}} \sqrt{g_{\alpha \beta}(\bs q, \dot{\bs q}) \dot{q}^{\alpha} \dot{q}^{\beta}} \\
&= \frac{1}{2} \frac{1}{\Lambda (\bs q, \dot{\bs q})} \frac{\partial g_{\alpha \beta}(\bs q, \dot{\bs q})}{\partial q^{\gamma}}\dot{q}^{\alpha} \dot{q}^{\beta}\\
&=  \frac{1}{2}  \frac{\partial g_{\alpha \beta}(\bs q, \dot{\bs q})}{\partial q^{\gamma}}\dot{q}^{\alpha} \dot{q}^{\beta} \, .
\label{eq_lhs}
\end{align} 
The last equality is valid when we choose the arc-length parameter $s$ for $\tau$, 
so that, by definition, $\Lambda (\bs q, \dot{\bs q})=1$ along the path.\\

We can therefore rewrite Eq.~\eqref{eq_rhs} as follows
\begin{equation}
 \frac{d}{d s} (g_{\gamma \beta} (\bs q, \dot{\bs q}) \dot{q}^{\beta}) - \frac{1}{2}  \frac{\partial g_{\alpha \beta}(\bs q, \dot{\bs q})}{\partial q^{\gamma}}\dot{q}^{\alpha} \dot{q}^{\beta} =0 \, ,
\label{eq_rhs_lhs}
\end{equation}
which leads to
\begin{equation}
g_{\gamma \beta} (\bs q, \dot{\bs q}) \,  \ddot{q}^{\beta} + 
\left( \frac{\partial g_{\gamma \alpha}(\bs q, \dot{\bs q})}{\partial q^{\beta}} - \frac{1}{2}  \frac{\partial g_{\alpha \beta}(\bs q, \dot{\bs q})}{\partial q^{\gamma}} \right) \dot{q}^{\alpha} \dot{q}^{\beta} 
=0 \, .
\label{eq_rhs_lhs2}
\end{equation}
Introducing the Christoffel symbols
\begin{equation}
\Gamma_{\alpha \beta \gamma}(\bs q, \dot{\bs q})=
\frac{1}{2} 
\left(
\frac{\partial g_{\alpha \beta}(\bs q, \dot{\bs q})}{\partial q^{\gamma}}+
\frac{\partial g_{ \beta \gamma}(\bs q, \dot{\bs q})}{\partial q^{\alpha}}-
\frac{\partial g_{\gamma \alpha}(\bs q, \dot{\bs q})}{\partial q^{\beta}}
\right)
\end{equation}
and the corresponding Christoffel symbols of second kind
\begin{equation}
?\Gamma^{\alpha}^{}_{\beta \gamma}?(\bs q, \dot{\bs q})= g^{\nu \alpha}(\bs q, \dot{\bs q}) \Gamma_{\beta \nu \gamma}(\bs q, \dot{\bs q})
\end{equation}
we arrive to the geodetic equations
\begin{equation}
\ddot{q}^{\alpha}+?\Gamma^{\alpha}^{}_{\beta \gamma}?(\bs q,\dot{\bs q}) \dot{q}^{\beta} \dot{q}^{\gamma} =0 \, 
\label{eq_geodesic}
\end{equation}
where $g^{\nu \alpha}(\bs q, \dot{\bs q})$ is defined as
\begin{equation}
g_{\nu \beta}(\bs q, \dot{\bs q}) g^{\nu \alpha}(\bs q, \bs p) =\delta_{\beta}^{\alpha} \, ,
\label{eq_geodesic2}
\end{equation}
and $p_{\alpha}=g_{\alpha \beta}(\bs q, \dot{\bs q}) \dot{q}^{\beta}$ (note that $\dot{\bs q}$ is a contravariant vector while $\bs p$ is a covariant vector).

\bigskip

In the derivation of Eq.~\eqref{eq_geodesic} we make use of the symmetry properties of $g_{\alpha \beta}$, namely the fact that
$g_{\alpha \beta}$ are positively homogeneous functions of degree zero in the variables $\dot{q}^{\alpha}$ and symmetric in their indices. As a consequence, the tensor defined as (H. Rund, Ref.~\cite{Rund59}, Chapter 1, section 3, p 15)
\begin{equation}
\Xi_{\alpha \beta \gamma}(\bs q,\dot{\bs q}) = \frac{\partial g_{\alpha \beta}(\bs q,\dot{\bs q})}{\partial \dot{q}^{\gamma}}=
\frac{1}{2} \frac{\partial^3 \Lambda^2(\bs q,\dot{\bs q})}{\partial \dot{q}^{\alpha} \partial \dot{q}^{\beta} \partial \dot{q}^{\gamma}} 
\label{eq_symm}
\end{equation}
is positively homogeneous of degree $-1$ and 
is symmetric in all three indices. Therefore
\begin{equation}
\Gamma_{\alpha \beta \gamma}(\bs q, \dot{\bs q}) \dot{q}^{\beta} \dot{q}^{\gamma} = \frac{1}{2} \frac{\partial g_{\beta \gamma}(\bs q, \dot{\bs q})}{\partial q^{\alpha}} \dot{q}^{\beta} \dot{q}^{\gamma} \,. 
\end{equation}

\bigskip

\noindent \textit{Derivation of Eq.~\eqref{Eq:deriv1}}:\\
\begin{align}
\frac{1}{2} \frac{\partial^2 \Lambda^2(\bs q, \dot{\bs q})}{\partial \dot{q}^{\alpha} \partial \dot{q}^{\beta}} \dot{q}^{\beta} &= 
\frac{1}{2} \left( \frac{\partial}{\partial \dot{q}^{\alpha}} \left(\frac{\partial}{\partial \dot{q}^{\beta}} \Lambda^2(\bs q, \dot{\bs q}) \right) \right) \dot{q}^{\beta} \notag \\
&= \frac{\partial}{\partial \dot{q}^{\alpha}} \left(\Lambda(\bs q, \dot{\bs q})\frac{\partial}{\partial \dot{q}^{\beta}}  \Lambda(\bs q, \dot{\bs q}) \right)\dot{q}^{\beta} \notag \\
&= \left(\frac{\partial}{\partial \dot{q}^{\alpha}}  \Lambda(\bs q, \dot{ \bs q}) \right)  \left(\frac{\partial}{\partial \dot{q}^{\beta}}  \Lambda(\bs q, \dot{\bs q}) \right)  \dot{q}^{\beta} + \Lambda(\bs q, \dot{\bs q}) \left( \frac{\partial}{\partial \dot{q}^{\alpha}} \left(\frac{\partial}{\partial \dot{q}^{\beta}}  \Lambda(\bs q, \dot{\bs q})  \right) \right) \dot{q}^{\beta} \notag \\
&= \left(\frac{\partial}{\partial \dot{q}^{\alpha}}  \Lambda(\bs q, \dot{\bs q}) \right)  \left(\frac{\partial}{\partial \dot{q}^{\beta}}  \Lambda(\bs q, \dot{\bs q}) \right)  \dot{q}^{\beta} \label{eq:deriv2}\\
& = \Lambda(\bs q, \dot{\bs q}) \frac{\partial \Lambda(\bs q, \dot{\bs q})}{\partial \dot{q}^{\alpha}} 
\end{align}
Eq.~\eqref{eq:deriv2} derives from Euler's theorem on homogeneous functions (Eq.~\eqref{eq_euler2}), 
while the last equality is a consequence of the homogeneity of $\Lambda(\bs q, \dot{\bs q})$ with respect to $\dot{q}^{\alpha}$ (Eq.~\eqref{eq_euler1}).

\section{Addition of a time-independent classical potential to the geodesic dynamics}

The addition of a classical time-independent potential to the geodesic motion
\begin{equation}
\ddot{q}^{\alpha}+?\Gamma^{\alpha}^{}_{\beta \gamma}?(\bs q,\dot{\bs q}) \dot{q}^{\beta} \dot{q}^{\gamma} =0 \, \end{equation}
gives (\cite{Abraham_Marsden})
\begin{equation}
\ddot{q}^{\alpha}+?\Gamma^{\alpha}^{}_{\beta \gamma}?(\bs q,\dot{\bs q}) \dot{q}^{\beta} \dot{q}^{\gamma} + {g}^{\alpha \beta}\frac{\partial V(\{q\}_{i=1}^n)}{\partial q^\beta}=0 \, .
\label{eq_geodesic2}
\end{equation}

\section{Derivation of the metric tensor components}

In this section, I derive the components of the metric tensor $g_{\alpha \beta}(\bs q,\dot{\bs q})$.

\begin{itemize}
\item[a)]   $g_{00}(\bs q,\dot{\bs q}) = \frac{1}{2}\frac{\partial^2 \Lambda^2(\bs q,\dot{\bs q})}{\partial \dot{q}^{0} \partial \dot{q}^{0}} $.\\

From

\begin{itemize}
\item[] $\Lambda(\bs q,\dot{\bs q})= \mathcal{T} \frac{1}{\dot{q}^0} - Q(\bs q) \dot{q}^{0}$
\item[] $\Lambda^2(\bs q,\dot{\bs q})= (\mathcal{T} \frac{1}{\dot{q}^0} - Q(\bs q) \dot{q}^{0})^2$
\item[] $\frac{\partial \Lambda^2(\bs q,\dot{\bs q})}{ \partial \dot{q}^{0}}=2 \Lambda(\bs q,\dot{\bs q}) \Lambda'(\bs q,\dot{\bs q})$
\end{itemize}
where
\begin{itemize}
\item[] $\Lambda'(\bs q,\dot{\bs q}) = -\mathcal{T} \frac{1}{(\dot{q}^0)^2} - Q_0(\bs q,t) \dot{q}^0 - Q(\bs q,t)$
\item[] $\Lambda''(\bs q,\dot{\bs q}) = 2\mathcal{T} \frac{1}{(\dot{q}^0)^3} - Q_{00}(\bs q,t) \dot{q}^0 - 2 Q_0(\bs q,t)$
\item[] $\left(\Lambda'(\bs q,\dot{\bs q})\right)^2= \frac{\mathcal{T}^2}{(\dot{q}^0)^4} +\frac{2 \mathcal{T} Q_0(\bs q,t)}{\dot{q}^0} + 
2 \frac{\mathcal{T} Q(\bs q,t)}{(\dot{q}^0)^2} + 2 Q(\bs q,t) Q_0(\bs q,t) \dot{q}^0 + Q^2_0(\bs q,t) (\dot{q}^0)^2 + Q^2(\bs q,t)$
\item[] $\Lambda(\bs q,\dot{\bs q}) \Lambda''(\bs q,\dot{\bs q}) = \frac{2 \mathcal{T}^2}{(\dot{q}^0)^4} - \mathcal{T} Q_{00}(\bs q,t) 
-\frac{2 \mathcal{T} Q_0(\bs q,t)}{\dot{q}^0}  
- \frac{2 \mathcal{T} Q(\bs q,t)}{(\dot{q}^0)^2} 
+ Q(\bs q,t) Q_{00}(\bs q,t) (\dot{q}^0)^2
+ 2 Q(\bs q,t) Q_0(\bs q,t) \dot{q}^0$
\end{itemize}

and $\mathcal{T}=\frac{1}{2} \sum_i m_i (\dot{q}^i)^2$, 
$Q_0(\bs q,t)=\frac{\partial Q(\bs q,t)}{\partial \dot{q}^0}$, and 
$Q_{00}(\bs q,t)=\frac{\partial^2 Q(\bs q,t)}{\partial \dot{q}^0\partial \dot{q}^0 }$\\

we get
\begin{equation}
\frac{1}{2} \frac{\partial^2 \Lambda^2(\bs q,\dot{\bs q})}{\partial \dot{q}^{0} \partial \dot{q}^{0}} = 
\left(\Lambda'(\bs q,\dot{\bs q})\right)^2 + \Lambda(\bs q,\dot{\bs q}) \Lambda''(\bs q,\dot{\bs q})   \, .
\end{equation}

\item[b)]  $g_{0i}(\bs q,\dot{\bs q}) = \frac{1}{2}\frac{\partial^2 \Lambda^2(\bs q,\dot{\bs q})}{\partial \dot{q}^{i} \partial \dot{q}^{0}} $.\\

\begin{align}
\frac{1}{2} \frac{\partial^2 \Lambda^2(\bs q,\dot{\bs q})}{\partial \dot{q}^{i} \partial \dot{q}^{0}} &= 
 \frac{\partial}{\partial \dot{q}^i}  \Lambda(\bs q,\dot{\bs q}) \Lambda'(\bs q,\dot{\bs q}) \\
 &= \frac{\partial}{\partial \dot{q}^i} \left(\frac{\mathcal{T}}{\dot{q}^0} -Q(\bs q,t) \dot{q}^0 \right) 
 \left(-\frac{\mathcal{T}}{(\dot{q}^0)^2} -Q_0(\bs q,t) \dot{q}^0 -Q(\bs q,t)\right)\\
 &= -\frac{2 \mathcal{T}(\dot{\bs q}) \mathcal{T}_i(\dot{\bs q})}{(\dot{q}^0)^3} - Q_0(\bs q,t) \mathcal{T}_i(\dot{\bs q}) 
\end{align}
where $\mathcal{T}_i(\dot{\bs q})=\frac{\partial \mathcal{T}(\dot{\bs q})}{\partial \dot{q}^i}$.

\item[c)] $g_{ij}(\bs q,\dot{\bs q})= \frac{1}{2}\frac{\partial^2 \Lambda^2(\bs q,\dot{\bs q})}{\partial \dot{q}^{i} \partial \dot{q}^{j}} $.\\

\begin{align}
\frac{1}{2} \frac{\partial^2 \Lambda^2(\bs q,\dot{\bs q})}{\partial \dot{q}^{i} \partial \dot{q}^{j}} &= 
\frac{1}{2} \frac{\partial}{\partial \dot{q}^j} \left(2 \Lambda(\bs q,\dot{\bs q}) \frac{\Lambda(\bs q,\dot{\bs q})}{\partial \dot{q}^i}\right) \\
&= \left (m_i \delta_{il} \frac{\dot{q}^l}{\dot{q}^0} \right) \left (m_j \delta_{jk} \frac{\dot{q}^k}{\dot{q}^0} \right) + \Lambda(\bs q,\dot{\bs q}) \delta_{ij} m_i \frac{1}{\dot{q}^0} \, .
\end{align}

\end{itemize}

Summarizing, the components of the metric tensor $g_{\alpha \beta}(\bs q,\dot{\bs q})$ are
\begin{align}
g_{00}(\bs q,\dot{\bs q})&= 3 \frac{\mathcal{T}^2}{(\dot{q}^0)^4} + Q^2(\bs q,t) + 4 Q(\bs q,t) {Q_{0}}(\bs q,t) \dot{q}^0 + {Q^2_{0}}(\bs q,t) (\dot{q}^0)^2 +Q^2_{0}(\bs q,t) (\dot{q}^0)^2  \notag \\ 
&\quad  - \mathcal{T} Q_{00}(\bs q,t) +Q(\bs q,t) {Q_{00}}(\bs q,t) (\dot{q}^0)^2 \label{eq:g00}\\
g_{0i}(\bs q,\dot{\bs q})&= - (m_j \dot{q}^j) \left( 2  \frac{\mathcal{T}}{(\dot{q}^0)^3}   + {Q_0}(\bs q,t) \right) \label{eq:g0i}\\
g_{ij}(\bs q,\dot{\bs q})&=(m_i \frac{\dot{q}^i}{\dot{q}^0}) (m_j \frac{\dot{q}^j}{\dot{q}^0}) + \left(\frac{\mathcal{T}}{(\dot{q}^0)^2} -Q(\bs q,t)\right) m_i \delta_{ij} \label{eq:gij} \, .
\end{align}

\section{Derivation of Eqs.12-13}

Using the polar representation of the system wavefunction
\begin{equation} 
\phi(\bs x,t)=A(\bs x,t) e^{i S(\bs x,t)/\hbar} \label{eq:polar}
\end{equation}
in the Lagrangian density $\mathscr{L}$ (Eq.(5)), 
we get ($\rho(\bs x,t)=A^2(\bs x,t)$)
\begin{equation}
\mathscr{L}(\rho, \{\partial_i \rho\}_{i=1}^n,\dot{\rho}, S, \{\partial_i S\}_{i=1}^n,\dot{S})=
-\rho(\bs x,t) 
\left[
\dot{S}(\bs x,t) +\frac{(\nabla S(\bs x,t))^2}{2m} +V(\bs x)
\right]
- \frac{\hbar^2 (\nabla \rho(\bs x,t))^2}{8 m \rho(\bs x,t)} \, .
\end{equation}
Following~\cite{hollandbook}, the field equation (Eq.(6)) can be written in the form of an energy transport equation
\begin{equation} \label{eq:tr1}
\frac{?T^0^{}_0?(\bs x,t)}{\partial t} + ?T^{k}^{}_{0,k}?(\bs x,t) = - \rho \frac{\partial V(\bs x)}{\partial t} \, ,
\end{equation}
which is equivalent to the quantum Hamilton-Jacobi field and continuity equations (the proof is given in Appendix E)
\begin{align}
&\frac{\partial S(\bs x,t)}{\partial t} + \frac{(\nabla S(\bs x,t))^2}{2 m} - \frac{\hbar^2}{2m} \frac{\nabla^2 A(\bs x,t)}{A(\bs x,t)} + V(\bs x)  = 0 \label{HJ1}\\
&\frac{\partial A^2(\bs x,t)}{\partial t} + \nabla \cdot \left( \frac{A^2(\bs x,t) \nabla S(\bs x,t)}{m} \right) = 0 \label{HJ2}\, .
\end{align}
In Eq.~\eqref{eq:tr1} we use the notation $?T^{k}^{}_{i,k}?(\bs x,t)=\partial ?T^{k}^{}_{i}?(\bs x,t)/ \partial x_k$.
Note that Eqs.~\eqref{HJ1} and~\eqref{HJ2} can also be derived directly from the time-dependent Schr\"odinger equation
\begin{equation}
i\hbar \frac{\partial \phi(\bs x,t)}{\partial t}=\left(-\frac{\hbar^2}{2 m} + V(\bs x) \right) \phi(\bs x,t)
\end{equation}
using the polar expansion in Eq.~\eqref{eq:polar}.

Eq.~\eqref{HJ1} has the form of a classical Hamilton-Jacobi field equation apart for the extra term
\begin{equation}
Q(\bs x,t)=- \frac{\hbar^2}{2m} \frac{\nabla^2 A(\bs x,t)}{A(\bs x,t)} \, ,
\end{equation}
which is called the \textit{quantum potential}.
The system trajectory is defined by the equation $ \dot{p}= \nabla S(\bs x,t)/m$. Applying the operator $\nabla$ to Eq.~\eqref{HJ1} we get the following field equation
\begin{equation}
\left[\frac{\partial}{\partial t} + \frac{1}{m} \nabla S(\bs x,t) \cdot \nabla \right] \, \nabla S(\bs x,t) = - \nabla \left( V(\bs x)+Q(\bs x,t) \right) \, .
\end{equation}
Setting the particle velocity equal to $\nabla S(\bs x,t)/m$ and moving to the Lagrangian frame, we finally get
\begin{equation}
m \frac{d^2}{dt^2} \bs x(t)= - \nabla \left(V(\bs x)+Q(\bs x,t) \right)
\label{Bohm_traj1}
\end{equation}
with $d/dt = \partial/\partial t + \dot{\bs x} \cdot \nabla$.

Equivalently, Bohmian trajectories can be obtained from the solution of a first order differential equation
\begin{equation}
\frac{d}{dt} \bs x(t)= v^{\phi} (\bs x,t)
\label{Bohm_traj2}
\end{equation}
where $v^{\phi} (\bs x,t)$ is defined in Eq.~\eqref{theo1} of the main text. 
Taking the time derivative of Eq.~\eqref{Bohm_traj2} one recovers Eq.~\eqref{Bohm_traj1}.

\section{Proof of the equivalence of Eq.~\eqref{eq:tr1}  with Eqs.~\eqref{HJ1} and \eqref{HJ2} 
(or equivalenty of Eq.~\eqref{eq:field1.1} with Eqs.~\eqref{prop3} and \eqref{prop4})}

Following~\cite{hollandbook}, Eqs.~\eqref{HJ1} and~\eqref{HJ2} can be derived from
\begin{equation} 
\frac{?T^0^{}_0?(\bs x,t)}{\partial t} + ?T^{k}^{}_{0,k}?(\bs x,t) = - \rho \frac{\partial V(\bs x)}{\partial t}
\end{equation}
using Eq.~\eqref{eq:polar}.

The first term on the LHS becomes (we use the notation $\dot{F}=\frac{\partial F}{\partial t}$, and ${F}_k=\frac{\partial F}{\partial x_k}$ for any function $F(\bs x, t)$),
\begin{align}
\frac{\partial ?T^0^{}_0?(\bs x,t)}{\partial t} 
&=\partial_t \rho
\left[ 
\frac{(\nabla S)^2}{2m} + V
\right] + 
\rho \left[ 
{\partial_t} \frac{(\nabla S)^2}{2m} + {\partial_t} V
\right] + \frac{\hbar^2}{8 m} \partial_t \frac{(\nabla \rho)^2}{\rho} \notag \\
&= \dot{\rho}
\left[ 
\frac{(\nabla S)^2}{2m} + V
\right] + 
\rho \left[ 
\frac{2 \nabla S \nabla \dot{S}}{2m} + {\partial_t} V
\right] + \frac{\hbar^2}{8 m} \frac{2 (\nabla \rho) (\nabla \dot{\rho}) \rho - (\nabla \rho)^2 \dot{\rho}}{\rho^2} \notag \\
&= \dot{\rho}
\left[ 
\frac{(\nabla S)^2}{2m} + V
\right] + 
\rho \left[ 
\frac{  S_k  \dot{S}_k}{m} + {\partial_t} V
\right] + \frac{\hbar^2}{8 m} \frac{4 (\nabla A) A (\partial_t \nabla \rho) A^2- (2 A \nabla A)^2 2 A \dot{A}}{A^4} \notag \\
&= \dot{\rho}
\left[ 
\frac{(\nabla S)^2}{2m} + V
\right] + 
\rho \left[ 
\frac{  S_k  \dot{S}_k}{m} + {\partial_t} V
\right] + 
\frac{\hbar^2}{m} \frac{(\nabla A)^2 \dot{A}}{A} +  
\frac{\hbar^2}{m} A_k \dot{A}_k
-\frac{\hbar^2}{m} \frac{( \nabla A)^2  \dot{A}}{A} \notag \\
&= \dot{\rho}
\left[ 
\frac{(\nabla S)^2}{2m} + V
\right] + 
\rho \left[ 
\frac{  S_k  \dot{S}_k}{m} + {\partial_t} V
\right] +  
\frac{\hbar^2}{m} A_k \dot{A}_k \label{eq:t00}
\end{align}

The second term on the LHS becomes
\begin{align}
?T^{k}^{}_{0,k}?(\bs x,t) 
&= -\partial_k \left(\rho \dot{S} \partial_k S/m + \hbar^2 \frac{\dot{\rho}\partial_k \rho}{4m \rho} \right) \notag \\
&= - (\partial_k \rho) \dot{S} \frac{\partial_k S}{m} -\rho \frac{\partial_k(\dot{S} \partial_k S)}{m} -
\hbar^2 \left[
\partial_k \dot{\rho} \frac{\partial_k \rho}{4 m \rho} + \dot{\rho} 
\left(\frac{\partial_k^2 \rho}{4 m \rho} - \frac{(\partial_k \rho)^2 }{4 m \rho^2} \right)
\right] \notag \\
&= - \frac{2 A A_k \dot{S} S_k}{m} -A^2 \frac{\partial_k(\dot{S} S_k)}{m} -
\hbar^2 \left[
 \frac{(\partial_t \partial_k \rho) \partial_k \rho}{4 m \rho} + 2 A \dot{A} 
\left(\frac{1}{2 m} \left(\frac{(\nabla A)^2}{A^2}+\frac{\nabla^2A}{A}\right) - \frac{(2 A \nabla A)^2 }{4 m A^4} \right)
\right] \notag \\
&= - \frac{2 A A_k \dot{S} S_k}{m} -A^2 \frac{\partial_k(\dot{S} S_k)}{m} -
\frac{\hbar^2}{m} \left[
 \frac{(\partial_t  (A \partial_k A))  A \partial_k A}{ A^2} +  A \dot{A} 
\left(\left(\frac{(\nabla A)^2}{A^2}+\frac{\nabla^2A}{A}\right) - \frac{2 (A \nabla A)^2 }{ A^4} \right)
\right]  \notag \\
&= - \frac{2 A A_k \dot{S} S_k}{m} -A^2 \frac{\partial_k(\dot{S} S_k)}{m} -
\frac{\hbar^2}{m} \left[
 \frac{2 \dot{A} (\nabla A)^2}{A} + \partial_k \dot{A}  \partial_k A +   \dot{A} 
\left(\nabla^2A - \frac{2 ( \nabla A)^2 }{ A} \right)
\right] \notag  \\
&= - \frac{2 A A_k \dot{S} S_k}{m} 
-A^2 \frac{\dot{S}_ k S_k+ \dot{S} S_{kk}}{m} -
\frac{\hbar^2}{m} \left[
\dot{A}_k  A_k +   \dot{A} 
\left(\nabla^2A \right)
\right] \label{eq:tk0k}
\end{align}

Combining all terms (Eq.~\eqref{eq:t00} and~\eqref{eq:tk0k}), we get
\begin{align*}
&\frac{\partial ?T^0^{}_0?(\bs x,t)}{\partial t} + ?T^{k}^{}_{0,k}?(\bs x,t) = - \rho \frac{\partial V(\bs x)}{\partial t} \\
&\dot{\rho}
\left[ 
\frac{(\nabla S)^2}{2m} + V
\right] + 
\rho \left[ 
\frac{  S_k  \dot{S}_k}{m} + {\partial_t} V
\right] +  
\frac{\hbar^2}{m} A_k \dot{A}_k - \frac{2 A A_k \dot{S} S_k}{m} 
-A^2 \frac{\dot{S}_ k S_k+ \dot{S} \nabla^2 S}{m} - 
\frac{\hbar^2}{m} \left[
\dot{A}_k  A_k +   \dot{A} 
\left(\nabla^2A \right)
\right] =  \rho \frac{\partial V}{\partial t}
\\
&\dot{\rho}
\left[ 
\frac{(\nabla S)^2}{2m} + V
\right]  
 - \frac{2 A A_k \dot{S} S_k}{m} 
-\frac{A^2}{m} \dot{S} \nabla^2 S - 
\frac{\hbar^2}{m} \left[
\dot{A} 
\left(\nabla^2A \right)
\right] =  0
\\
&
\frac{1}{\dot{S}} \left[ 
\frac{(\nabla S)^2}{2m} + V
\right]  
-\frac{1}{\dot{A}} \frac{ A_k S_k}{ m} 
-\frac{1}{\dot{A}} \frac{A \nabla^2 S}{2 m}  -  
\frac{1}{\dot{S}} 
\frac{\hbar^2}{2m} 
\frac{\left(\nabla^2A \right)}{A}
=  0
\end{align*}

\noindent Symbolically, a system of differential equations like
\begin{align*}
\dot{S}&=f(S,A)\\
\dot{A}&=g(S,A)
\end{align*}
is equivalent to
\begin{equation*}
\frac{f(S,A)}{\dot{S}} - \frac{g(S,A)}{\dot{A}} = 0 \, .
\end{equation*}

We finally get that the field equation
\begin{equation*} 
\frac{\partial ?T^0^{}_0?(\bs x,t)}{\partial t} + ?T^{k}^{}_{0,k}?(\bs x,t) = - \rho \frac{\partial V(\bs x)}{\partial t}
\end{equation*}
can be split into the coupled differential equations
\begin{align}
&\frac{\partial S}{\partial t} +\frac{(\nabla S)^2}{2m} -\frac{\hbar^2}{2m} \frac{\nabla^2 A}{A} + V=0 \\
&\frac{\partial A^2}{\partial t} + \nabla \left( \frac{A^2 \nabla S}{m} \right)= 0 \, .
\end{align}


\begin{thebibliography}{34}
\expandafter\ifx\csname natexlab\endcsname\relax\def\natexlab#1{#1}\fi
\expandafter\ifx\csname bibnamefont\endcsname\relax
  \def\bibnamefont#1{#1}\fi
\expandafter\ifx\csname bibfnamefont\endcsname\relax
  \def\bibfnamefont#1{#1}\fi
\expandafter\ifx\csname citenamefont\endcsname\relax
  \def\citenamefont#1{#1}\fi
\expandafter\ifx\csname url\endcsname\relax
  \def\url#1{\texttt{#1}}\fi
\expandafter\ifx\csname urlprefix\endcsname\relax\def\urlprefix{URL }\fi
\providecommand{\bibinfo}[2]{#2}
\providecommand{\eprint}[2][]{\url{#2}}

\bibitem[{\citenamefont{Feynman et~al.}(2002)\citenamefont{Feynman, Morinigo,
  Wagner, and Hatfield}}]{Feynman_grav}
\bibinfo{author}{\bibfnamefont{R.}~\bibnamefont{Feynman}},
  \bibinfo{author}{\bibfnamefont{F.}~\bibnamefont{Morinigo}},
  \bibinfo{author}{\bibfnamefont{W.}~\bibnamefont{Wagner}}, \bibnamefont{and}
  \bibinfo{author}{\bibfnamefont{B.}~\bibnamefont{Hatfield}},
  \emph{\bibinfo{title}{Feynman Lectures on Gravitation}}
  (\bibinfo{publisher}{Westview Press}, \bibinfo{year}{2002}).

\bibitem[{com({\natexlab{a}})}]{comment_1}
\bibinfo{note}{It is important to distinguish between the mathematical
  formalisms of QM from their conceptual interpretations. The mathematical
  predictions of QM are well known and tested experimentally, independently
  from the formalism used: Schr\"odinger or Heisenberg formulations, particles
  or fields quantization, Bohm-trajectories or wavefunction-based descriptions.
  When properly interpreted, all these formalisms show a very high level of
  agreement with experiments. On the other hand, the conceptual interpretation
  of QM, i.e. the understanding of the physical/philosophical interpretations
  of QM phenomena like entanglement, non-locality, causality, and the
  appearance of the classical world is still the subject of a long-lasting
  debate. Different mathematical formulations require different interpretation
  schemes ranging from the `reduction of the wave packet' (Copenhagen
  interpretation) in the Schr\"odinger wavefucntion formalism, through the
  many-worlds interpretation, the hidden variable and pilot-wave models of de
  Broglie-Bohm theory, to the stochastic (quantum foam) interpretation, and
  many others. The focus of this article is to present a new geometrical
  framework for the description of quantum dynamics together with a suitable
  interpretative picture (geodesics in a curved space), which makes it
  consistent with all known experiments. It is beyond the scope of this work to
  investigate all conceptual implications of this and other formulations of
  QM.}

\bibitem[{\citenamefont{de~Broglie}(1926)}]{deBroglie26b}
\bibinfo{author}{\bibfnamefont{L.}~\bibnamefont{de~Broglie}},
  \bibinfo{journal}{Nature} \textbf{\bibinfo{volume}{118}}, \bibinfo{pages}{441}
  (\bibinfo{year}{1926}).

\bibitem[{\citenamefont{Bohm}(1952{\natexlab{a}})}]{Bohm1952a}
\bibinfo{author}{\bibfnamefont{D.}~\bibnamefont{Bohm}}, \bibinfo{journal}{Phys.
  Rev.} \textbf{\bibinfo{volume}{85}}, \bibinfo{pages}{166}
  (\bibinfo{year}{1952}{\natexlab{a}}).

\bibitem[{\citenamefont{Bohm}(1952{\natexlab{b}})}]{Bohm1952}
\bibinfo{author}{\bibfnamefont{D.}~\bibnamefont{Bohm}}, \bibinfo{journal}{Phys.
  Rev.} \textbf{\bibinfo{volume}{85}}, \bibinfo{pages}{180}
  (\bibinfo{year}{1952}{\natexlab{b}}).

\bibitem[{end({\natexlab{a}})}]{endnote_nonlocality}
\bibinfo{note}{Quantum mechanics is a nonlocal theory in the sense that every
  part of the quantum system depends on every other part even in the absence of
  an interacting potential, $V(\mathbf{x})=0$.}

\bibitem[{end({\natexlab{b}})}]{endnote1}
\bibinfo{note}{Bohmian dynamics can also be formulated as a first-oder equation
  of motion, which is mathematically and computationally more convenient, but
  less useful for a comparison with classical mechanics.}

\bibitem[{\citenamefont{Holland}(1993)}]{hollandbook}
\bibinfo{author}{\bibfnamefont{P.~R.} \bibnamefont{Holland}},
  \emph{\bibinfo{title}{The Quantum Theory of Motion - An Account of the de
  Broglie-Bohm Causal Interpretation of Quantum Mechanics}}
  (\bibinfo{publisher}{Cambridge University Press}, \bibinfo{year}{1993}).

\bibitem[{\citenamefont{D\"urr et~al.}(1992)\citenamefont{D\"urr, Goldstein,
  and Zanghi}}]{duerr92}
\bibinfo{author}{\bibfnamefont{D.}~\bibnamefont{D\"urr}},
  \bibinfo{author}{\bibfnamefont{S.}~\bibnamefont{Goldstein}},
  \bibnamefont{and} \bibinfo{author}{\bibfnamefont{N.}~\bibnamefont{Zanghi}},
  \bibinfo{journal}{J. Stat. Phys.} \textbf{\bibinfo{volume}{67}},
  \bibinfo{pages}{843} (\bibinfo{year}{1992}).

\bibitem[{\citenamefont{Rund}(1959)}]{Rund59}
\bibinfo{author}{\bibfnamefont{H.}~\bibnamefont{Rund}},
  \emph{\bibinfo{title}{The Differential Geometry of Finsler Spaces}}
  (\bibinfo{publisher}{Springer Verlag}, \bibinfo{year}{1959}).

\bibitem[{\citenamefont{Weyl}(1918)}]{weyl1918}
\bibinfo{author}{\bibfnamefont{H.}~\bibnamefont{Weyl}}, \bibinfo{journal}{Sitz.
  Ber. Preuss. Akad. Wiss. (Berlin)} pp. \bibinfo{pages}{466--480}
  (\bibinfo{year}{1918}).

\bibitem[{\citenamefont{Novello et~al.}(2011)\citenamefont{Novello, Salim, and
  Falciano}}]{novello2011}
\bibinfo{author}{\bibfnamefont{M.}~\bibnamefont{Novello}},
  \bibinfo{author}{\bibfnamefont{J.~M.} \bibnamefont{Salim}}, \bibnamefont{and}
  \bibinfo{author}{\bibfnamefont{F.~T.} \bibnamefont{Falciano}},
  \bibinfo{journal}{Int. J. Geom. Meth. Mod. Phys.}
  \textbf{\bibinfo{volume}{8}}, \bibinfo{pages}{87} (\bibinfo{year}{2011}).

\bibitem[{\citenamefont{Ootsuka and E.}(2010)}]{ootsuka2010}
\bibinfo{author}{\bibfnamefont{T.}~\bibnamefont{Ootsuka}} \bibnamefont{and}
  \bibinfo{author}{\bibfnamefont{E.}~\bibnamefont{Tanaka.}}, \bibinfo{journal}{Phys.
  Lett. A} \textbf{\bibinfo{volume}{371}}, \bibinfo{pages}{1917}
  (\bibinfo{year}{2010}).

\bibitem[{\citenamefont{Schouten}(1989)}]{Schouten}
\bibinfo{author}{\bibfnamefont{J.~A.} \bibnamefont{Schouten}},
  \emph{\bibinfo{title}{Tensor Analysis for Physicists}}
  (\bibinfo{publisher}{Dover Publications}, \bibinfo{year}{1989}).

\bibitem[{end({\natexlab{c}})}]{endnote0}
\bibinfo{note}{This is an important prerequisite for the geometrization
  process, which requires that all regular transformations (with non-zero
  determinant) of the variables defining the kinetic and the potential terms of
  the Lagrangian (hence including time in the case of a time-dependent
  potential) leave the Lagrange equation form-invariant.
  In general, the coordinates $x_i(t)$
  ($i=1,\dots,n$) are \textit{scalar} functions in $t$-space, in the sense that
  they do not undergo a transformation when time is scaled. One can
  show~\cite{Schouten} that in the extended configuration space the Lagrange
  equations \begin{equation*} \frac{d}{d\tau} \frac{\partial
  \Lambda(\bs q,\dot{\bs q})}{\partial \dot{q}^{\alpha}} -
  \frac{\partial \Lambda(\bs q,\dot{\bs q})}{\partial q^{\alpha}} =0
  \end{equation*} ($\alpha=0,\dots,n$) are invariant with respect to all
  regular transformations in the ($n+1$) variables $q_0=t, q_1, \dots q_{n}$
  and in the parameter $\tau$.}

\bibitem[{\citenamefont{Abraham and Marsden}(1994)}]{Abraham_Marsden}
\bibinfo{author}{\bibfnamefont{R.}~\bibnamefont{Abraham}} \bibnamefont{and}
  \bibinfo{author}{\bibfnamefont{J.~E.} \bibnamefont{Marsden}},
  \emph{\bibinfo{title}{Foundations of Mechanics}} (\bibinfo{publisher}{Addison
  Wesley}, \bibinfo{year}{1994}), chap. \bibinfo{chapter}{3.7: Mechanics on
  Riemannian Manifolds}.

\bibitem[{\citenamefont{Caratheodory}(1999)}]{Caratheodory}
\bibinfo{author}{\bibfnamefont{C.}~\bibnamefont{Caratheodory}},
  \emph{\bibinfo{title}{Calculus of Variations and Partial Differential
  Equations of First Order}} (\bibinfo{publisher}{American Mathematical
  Society}, \bibinfo{year}{1999}).

\bibitem[{\citenamefont{Schleich et~al.}(2013)\citenamefont{Schleich,
  Greenberger, Kobe, and Scully}}]{schleich2013schrodinger}
\bibinfo{author}{\bibfnamefont{W.~P.} \bibnamefont{Schleich}},
  \bibinfo{author}{\bibfnamefont{D.~M.} \bibnamefont{Greenberger}},
  \bibinfo{author}{\bibfnamefont{D.~H.} \bibnamefont{Kobe}}, \bibnamefont{and}
  \bibinfo{author}{\bibfnamefont{M.~O.} \bibnamefont{Scully}},
  \bibinfo{journal}{Proc. Natl. Acad. Sci. USA.}
  \textbf{\bibinfo{volume}{110}}, \bibinfo{pages}{5374} (\bibinfo{year}{2013}).

\bibitem[{\citenamefont{Takabayasi}(1952)}]{Takabayasi}
\bibinfo{author}{\bibfnamefont{T.}~\bibnamefont{Takabayasi}},
  \bibinfo{journal}{Prog. Theor. Phys.} \textbf{\bibinfo{volume}{8}},
  \bibinfo{pages}{143} (\bibinfo{year}{1952}).

\bibitem[{\citenamefont{Lurie}(1968)}]{Lurie}
\bibinfo{author}{\bibfnamefont{D.}~\bibnamefont{Lurie}},
  \emph{\bibinfo{title}{Particles and Fields}}
  (\bibinfo{publisher}{Interscience Publishers}, \bibinfo{year}{1968}),
  chap.~\bibinfo{chapter}{2}.

\bibitem[{\citenamefont{Misner et~al.}(1973)\citenamefont{Misner, Thorne, and
  Wheeler}}]{Misnerbook}
\bibinfo{author}{\bibfnamefont{C.}~\bibnamefont{Misner}},
  \bibinfo{author}{\bibfnamefont{K.~S.} \bibnamefont{Thorne}},
  \bibnamefont{and} \bibinfo{author}{\bibfnamefont{J.}~\bibnamefont{Wheeler}},
  \emph{\bibinfo{title}{Gravitation}} (\bibinfo{publisher}{Freeman, W.H. and
  Company}, \bibinfo{year}{1973}), chap. \bibinfo{chapter}{Stress-Energy Tensor
  and Conservation Laws}.

\bibitem[{\citenamefont{Curchod and Tavernelli}(2013)}]{nabdy2}
\bibinfo{author}{\bibfnamefont{B.~F.~E.} \bibnamefont{Curchod}}
  \bibnamefont{and}
  \bibinfo{author}{\bibfnamefont{I.}~\bibnamefont{Tavernelli}},
  \bibinfo{journal}{J. Chem. Phys.} \textbf{\bibinfo{volume}{138}},
  \bibinfo{pages}{184112} (\bibinfo{year}{2013}).

\bibitem[{\citenamefont{Curchod et~al.}(2013)\citenamefont{Curchod,
  Rothlisberger, and Tavernelli}}]{nabdy3}
\bibinfo{author}{\bibfnamefont{B.~F.~E.} \bibnamefont{Curchod}},
  \bibinfo{author}{\bibfnamefont{U.}~\bibnamefont{Rothlisberger}},
  \bibnamefont{and}
  \bibinfo{author}{\bibfnamefont{I.}~\bibnamefont{Tavernelli}},
  \bibinfo{journal}{ChemPhysChem} \textbf{\bibinfo{volume}{14}},
  \bibinfo{pages}{1314} (\bibinfo{year}{2013}).

\bibitem[{\citenamefont{Valentini}(1997)}]{valentini2008}
\bibinfo{author}{\bibfnamefont{A.}~\bibnamefont{Valentini}},
  \emph{\bibinfo{title}{On {Galilean} and {Lorentz} invariance in pilot-wave
  dynamics}}, \bibinfo{journal}{Phys. Lett. A} \textbf{\bibinfo{volume}{228}},
  \bibinfo{pages}{215} (\bibinfo{year}{1997}).

\bibitem[{com({\natexlab{b}})}]{comment_2}
\bibinfo{note}{In this geometrical formulation of quantum mechanics nonlocality
  arises from the propagation of the space curvature, which can diffuse (if
  coherence is maintained) at large distances from the `generating' particle.}

\bibitem[{com({\natexlab{c}})}]{comment_3}
\bibinfo{note}{Stationary states are described by equilibrium distributions of
  points particles for which the effect of the space curvature is exactly
  compensated by the Coulomb potential acting on the same particles.}

\bibitem[{\citenamefont{Kendrick}(2003)}]{kendrick2003new}
\bibinfo{author}{\bibfnamefont{B.~K.} \bibnamefont{Kendrick}},
  \bibinfo{journal}{J. Chem. Phys.} \textbf{\bibinfo{volume}{119}},
  \bibinfo{pages}{5805} (\bibinfo{year}{2003}).

\bibitem[{\citenamefont{Nairz et~al.}(2001)\citenamefont{Nairz, Brezger, Arndt,
  and Zeilinger}}]{PhysRevLett.87.160401}
\bibinfo{author}{\bibfnamefont{O.}~\bibnamefont{Nairz}},
  \bibinfo{author}{\bibfnamefont{B.}~\bibnamefont{Brezger}},
  \bibinfo{author}{\bibfnamefont{M.}~\bibnamefont{Arndt}}, \bibnamefont{and}
  \bibinfo{author}{\bibfnamefont{A.}~\bibnamefont{Zeilinger}},
  \bibinfo{journal}{Phys. Rev. Lett.} \textbf{\bibinfo{volume}{87}},
  \bibinfo{pages}{160401} (\bibinfo{year}{2001}).

\bibitem[{\citenamefont{Juffmann et~al.}(2009)\citenamefont{Juffmann, Truppe,
  Geyer, Major, Deachapunya, Ulbricht, and Arndt}}]{PhysRevLett.103.263601}
\bibinfo{author}{\bibfnamefont{T.}~\bibnamefont{Juffmann}},
  \bibinfo{author}{\bibfnamefont{S.}~\bibnamefont{Truppe}},
  \bibinfo{author}{\bibfnamefont{P.}~\bibnamefont{Geyer}},
  \bibinfo{author}{\bibfnamefont{A.~G.} \bibnamefont{Major}},
  \bibinfo{author}{\bibfnamefont{S.}~\bibnamefont{Deachapunya}},
  \bibinfo{author}{\bibfnamefont{H.}~\bibnamefont{Ulbricht}}, \bibnamefont{and}
  \bibinfo{author}{\bibfnamefont{M.}~\bibnamefont{Arndt}},
  \bibinfo{journal}{Phys. Rev. Lett.} \textbf{\bibinfo{volume}{103}},
  \bibinfo{pages}{263601} (\bibinfo{year}{2009}).

\bibitem[{\citenamefont{Juffmann et~al.}(2012)\citenamefont{Juffmann, Milic,
  M{\"u}llneritsch, Asenbaum, Tsukernik, T{\"u}xen, Mayor, Cheshnovsky, and
  Arndt}}]{juffmann2012real}
\bibinfo{author}{\bibfnamefont{T.}~\bibnamefont{Juffmann}},
  \bibinfo{author}{\bibfnamefont{A.}~\bibnamefont{Milic}},
  \bibinfo{author}{\bibfnamefont{M.}~\bibnamefont{M{\"u}llneritsch}},
  \bibinfo{author}{\bibfnamefont{P.}~\bibnamefont{Asenbaum}},
  \bibinfo{author}{\bibfnamefont{A.}~\bibnamefont{Tsukernik}},
  \bibinfo{author}{\bibfnamefont{J.}~\bibnamefont{T{\"u}xen}},
  \bibinfo{author}{\bibfnamefont{M.}~\bibnamefont{Mayor}},
  \bibinfo{author}{\bibfnamefont{O.}~\bibnamefont{Cheshnovsky}},
  \bibnamefont{and} \bibinfo{author}{\bibfnamefont{M.}~\bibnamefont{Arndt}},
  \bibinfo{journal}{Nature Nanotechnology} \textbf{\bibinfo{volume}{7}},
  \bibinfo{pages}{297} (\bibinfo{year}{2012}).

\bibitem[{\citenamefont{Philippidis et~al.}(1979)\citenamefont{Philippidis,
  Dewdney, and Hiley}}]{Philippidis1979}
\bibinfo{author}{\bibfnamefont{C.}~\bibnamefont{Philippidis}},
  \bibinfo{author}{\bibfnamefont{C.}~\bibnamefont{Dewdney}}, \bibnamefont{and}
  \bibinfo{author}{\bibfnamefont{B.~J.} \bibnamefont{Hiley}},
  \bibinfo{journal}{Il {Nuovo} {Cimento} B} \textbf{\bibinfo{volume}{52}},  \bibinfo{pages}{15}
  (\bibinfo{year}{1979}).

\bibitem[{\citenamefont{Rarity et~al.}(1990)\citenamefont{Rarity, Tapster,
  Jakeman, Larchuk, Campos, Teich, and B.E.A.}}]{Rarity1990}
\bibinfo{author}{\bibfnamefont{J.}~\bibnamefont{Rarity}},
  \bibinfo{author}{\bibfnamefont{P.}~\bibnamefont{Tapster}},
  \bibinfo{author}{\bibfnamefont{E.}~\bibnamefont{Jakeman}},
  \bibinfo{author}{\bibfnamefont{T.}~\bibnamefont{Larchuk}},
  \bibinfo{author}{\bibfnamefont{R.}~\bibnamefont{Campos}},
  \bibinfo{author}{\bibfnamefont{M.~C.} \bibnamefont{Teich}}, \bibnamefont{and}
  \bibinfo{author}{\bibfnamefont{S.}~\bibnamefont{B.E.A.}},
  \bibinfo{journal}{Phys. Rev. Lett.} \textbf{\bibinfo{volume}{65}},
  \bibinfo{pages}{1348} (\bibinfo{year}{1990}).

\bibitem[{end({\natexlab{d}})}]{endnote3}
\bibinfo{note}{In Bohmian dynamics the role of the space curvature is taken by
  what is usually referred to as the \textit{empty wave}, which takes the
  alternative path in the interferometer (the one not taken by the particle).}

\bibitem[{com({\natexlab{d}})}]{comment_4}
\bibinfo{note}{As in the case general relativity, each particle generates its
  own space curvature, which however does not give rise, \textit{per se}, to a
  dynamics. It is the curvature induced by the other particles in the system
  that determines the dynamics of the particle of interest along its geodesic.
  A `flat space' dynamics corresponds therefore to the time evolution of an
  isolated particle, which does not feel the curvature induced by the rest of
  the system.}

\end{thebibliography}

\end{document}